\documentclass[twocolumn,reprint,aip,apl,showpacs,superscriptaddress,groupedaddress]{revtex4-1}
\usepackage{amsmath,amssymb,color}
\usepackage{float}
\usepackage[caption=false]{subfig}
\usepackage{graphicx}
\usepackage{epstopdf}
\usepackage{natbib}
\begin{document}

\title{Large Magnetocaloric effect and magnetic phase transitions in Nd$_2$NiMnO$_6$}

\author{Gyaneswar Sharma, Kanika Pasrija, Anzar Ali, Sanjeev Kumar and Yogesh Singh}

\affiliation{%
Indian Institute of Science Education and Research Mohali, Sector 81, S.A.S. Nagar, Manauli PO 140306, India. \\
}

\date{\today}


\begin{abstract}
We present combined experimental and theoretical investigations on the magnetic and magnetocaloric behavior of Nd$_2$NiMnO$_6$.  The relative cooling power (RCP) which quantifies the usefulness of a magnetocaloric (MC) material is estimated to be $\approx 300$~J/Kg near the ferromagnetic transition at $T_C \approx 195$~K\@. This RCP is comparable to the best known MC materials.  Additionally, the magnetic entropy change has a broad profile ($T_C - 50~{\rm K} < T < T_C + 50~{\rm K}$) leading to an enhancement in the working-range of temperatures 
for magnetocaloric based cooling.  These features make Nd$_2$NiMnO$_6$ a superior magnetocaloric material compared for example, to the nonmagnetic counterpart
Y$_2$NiMnO$_6$.  We identify the mechanism for the enhanced RCP which can guide search for future MC materials.
\end{abstract}


\maketitle


{\it Introduction:}
Double perovskites (DPs) with the general formula {A$_2$BB}$^{'}${O$_6$}, where A is a rare-earth, 
alkali or alkaline-earth ion and B, B$'$ are transition metal ions, form a fundamentally interesting and technologically important class of oxides \cite{Vasala2015, Saha-Dasgupta2013, Aligia2001, Chattopadhyay2001, Sanyal2009, Kumar2010a, Chen2010a}.
DPs exhibit a variety of application oriented phenomena such as, magnetocaloric effect, magnetoelectric coupling, magnetocapacitance, 
magnetoresistance, magnetoelasticity and thermoelectric effect \cite{Yanez-Vilar2011, Rogado2005, Tokura1998, Lezaic2011, Singh2010, Sharma2013, Sharma2014, Serrate2007a, VillarArribi2016,Coy2016}.  The origin of most of these phenomena is tied to the existence of various magnetic phase transitions. 
 
In recent years, Ni-Mn and Co-Mn DPs have been investigated for the magnetocaloric effect (MCE) \cite{Sharma2013,Sharma2014,KrishnaMurthy2015}. {Y$_2$NiMnO$_6$} shows a second order
phase transition to ferromagnetic state at $T_{C} = 93$K, and a significant MCE was reported in the temperature range $T_{C}-10K < T < T_{C}+10K$. 
The refrigeration efficiency of a magnetocaloric material can be quantified in terms of relative cooling power (RCP).
RCP is defined as the product of full width at 
half maximum $(\Delta T_{{\rm FWHM} })$ and $\Delta S_{M}^{max}$ (maximum change in entropy) 
 \cite{Tegus2002,DeOliveira2008,K.A.Gschneidner2003}. In first order magnetic transitions, the $\Delta S_{M}^{max}$ is typically very large and hence leads to
a large RCP.
Indeed, giant magnetocaloric effect has been observed in materials such as {Gd$_5$Si$_{2}$Ge$_{2}$} and {MnAs} that undergo first-order 
phase transition near room temperature with RCP $\approx$ 278 J/Kg and 390 J/Kg, respectively \cite{Pecharsky1997,K.A.Gschneidner2003,Wada2001,Pecharsky1999}.  One potential drawback of materials having a first-order transition is that the extremely small $\Delta T_{{\rm FWHM} }$ makes the temperature range of their applicability rather small.  In a second-order phase transition while $\Delta S_{M}^{max}$ is suppressed $\Delta T_{{\rm FWHM} }$ can increase leading to an enhanced RCP.  For example, {Gd$_6$Co$_{1.67}$Si$_3$} has a much lower value of $\Delta S_{M}^{max}$ compared to
{Gd$_5$Si$_{2}$Ge$_{2}$} but still has an RCP of $\sim 310$J/Kg~  \cite{Jun2008}. 
With large ordering temperatures, second order magnetic phase transitions, and the possibility of tuning the size of magnetic moments residing at A and B/B' sites, DPs could turn out to be ideal candidates for large magnetocaloric effect. To the best of our knowledge, only a few explorations have been carried out in this direction \cite{Sharma2013, Sharma2014, KrishnaMurthy2015}.

In this work, we present a combined experimental and theoretical study of the MCE potential of Nd$_2$NiMnO$_6$. We find that Nd$_2$NiMnO$_6$ has a higher ferromagnetic ordering temperature ($193$K) compared to other DPs, studied so far, for the MCE.  A second order phase transition with a large $\Delta T_{{\rm FWHM} }$ leads to an RCP $\approx 300$ J/Kg that is comparable to the best known magnetocaloric materials. 
We present a theoretical analysis of the results using Monte Carlo simulations on a phenomenological magnetic model for Nd$_2$NiMnO$_6$. The presence of a paramagnetic background (of Nd$^{3+}$ ions) interacting with the ferromagnetically ordered ions leads to a broader $\Delta S_{M}$ profile and hence a larger RCP and an extended working temperature range. This allows us to propose certain general scenarios for the enhancement of the MCE.

{\it Experimental Results:} Polycrystalline Nd$_2$NiMnO$_6$ was synthesized using the conventional solid state reaction method as described elsewhere. The magnetization measurements were performed using the VSM option on a Quantum Design physical property measurement system.  Rietveld refinement of the room temperature powder X-ray pattern confirms that single phase samples crystallizing in the monoclinic structure with space group $P2_1/n$ have been obtained. The lattice parameters obtained from the refinement are $a=5.4150(5) \AA $, $b=5.4882(4) \AA $, $c=7.6770(6) \AA $, $ \alpha =90^\circ$, $\beta= 90.136(7)^\circ$, and $\gamma=90^\circ$. The lattice parameters match well with literature \cite{Sazonov2007}.  

 \begin{figure}[!t]
 \centering
 \includegraphics[width=1.0\columnwidth,angle=0, clip = 'True']{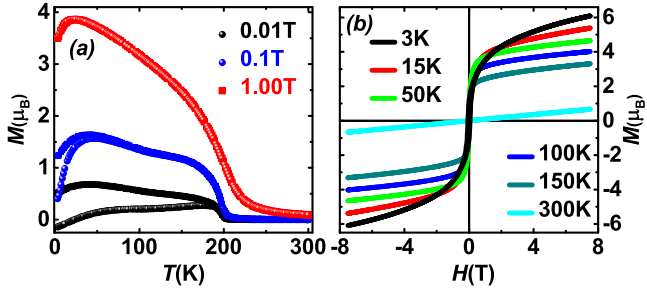}
 \caption{(Color online) (a) FC and ZFC magnetization curves recorded under magnetic field ($H$) of strength of 0.01 T, 0.1 T and 1.00 T. Splitting of ZFC and FC branch for $H=0.01$ T starts below ferromagnetic transition temperature, $T_{\mathrm C}$. (b) Isothermal curves of magnetization verses magnetic field are observed at various temperatures indicated in the plot.}
 \label{fig43}
 \end{figure} 
 
Fig. \ref{fig43} ($a$) shows magnetization versus temperature, $M(T)$, measured under zero-field cooled (ZFC) and field cooled (FC) protocol in an applied magnetic field $H = 0.01, 0.1,$ and $1$~T\@. The sudden increase in magnetization below $T_c \approx 195$~K signals the onset of the ferromagnetic transition consistent with a previous report \cite{Asai1998}.  The splitting of ZFC and FC $M(T)$ data below $T_c$ is also consistent with a ferromagnetic transition. The splitting decreases with increasing $H$ and is completely suppressed at $H = 1$ T\@.  A downturn in $M(T)$ data at $H = 0.01$~T below $\approx 50$~K is suggestive of antiferromagnetic correlations developing below this temperature. These results suggest that there are weaker antiferromagnetic interactions in the system in addition to the dominant ferromagnetic interactions.  Inverse susceptibility $1/\chi = H/M$ versus $T$ data above $T = 250$~K were fit by the Curie-Weiss expression, $\chi = C/(T-\theta)$, where $C$ is the Curie constant and $\theta$ is the Weiss temperature.  The fit gave the values $C = 5.18$~cm$^3$~K/mol and $\theta = 200$~K\@.  From $C$ we estimate the effective magnetic moment $\mu_{eff} = 6.44~\mu_B$ which is close to the expected value $\mu = 7~\mu_B$ \cite{}.

Figure~\ref{fig43}~($b$) shows the magnetization versus field $M(H)$ data recorded at various temperatures $T$. In the paramagnetic (PM) state at $T = 300$~K, $M(H)$ varies linearly and below the ferromagnetic critical temperature $T_{\mathrm{c}} \approx 195$~K, $M(H)$ shows rapid saturation accompanied by a small coercivity.  The $M(H)$ however, never saturates even at our highest applied magnetic fields, stays smaller than the expected saturation value $M_S = 8.3 \mu_B$, and there is a paramagnetic (approximately linear) component at all measured temperatures.  This suggests that only some of the magnetic ions participate in the ferromagnetic ordering and the rest stay paramagnetic.  This as we will see later, is an important ingredient for the enhanced MC behaviour. 

 \begin{figure}[!b]
 \centering
 \includegraphics[width=1.0\columnwidth,angle=0, clip = 'True']{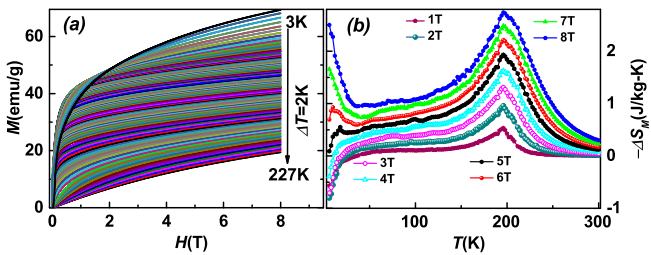}
 \caption{(Color online) (a) Series of isotherm of magnetization from 3 K to 227 K with a step size $\Delta T=2$ K. (b) Thermal profile of field induced change of magnetic entropy, $\Delta S_M$, calculated from Maxwell's equation (\ref{Maxwell_eqn}) using isothermal magnetization curve from 3 K to 301 K, under various external magnetic fields.}
 \label{fig44}
 \end{figure}
 
In order to understand the field dependent magnetic behavior of {Nd$_{2}$NiMnO$_{6}$} over a broad temperature range and to determine its magnetocaloric potential, $M(H)$ at various temperatures between $T = 2$~K and $305$~K were collected and are shown in Fig.~\ref{fig44}~($a$). Field and temperature dependent change in magnetic entropy (-$\Delta S_M$) were then extracted by integrating the area between various isotherms by applying Maxwell's thermodynamical relation as stated below \cite{Franco2012} 
\begin{equation}
 \Delta S_M (H,T)=\int_{0}^{H} \left( \dfrac{dM}{dT}\right) _{H'} dH'. \label{Maxwell_eqn}
\end{equation}
   
\noindent
The temperature dependence of -$\Delta S_M$ thus obtained for various magnetic fields are shown in Fig.~\ref{fig44}~($b$).  There are two features of interest near $T_c$ and near $T \approx 50$~K, respectively.  The temperature profile of -$\Delta S_M$ peaks smoothly near $T_C$ and the magnitude of the peak increases with increasing magnetic field reaching a maximum of -$\Delta S_M \approx 3$~J/Kg~K at $H = 8$~T\@. A -$\Delta S_M>0$ is characteristic of a ferromagnetic ordering \cite{Sharma2013}.  The temperature range for significant magnetocaloric effect can be estimated from the full-width at half maximum (FWHM) of -$\Delta S_M$. It is found that $\Delta T_{{\rm FWHM}} \approx 100$ K for an applied field of $8$~T\@.  The relative cooling power can be estimated as RCP = $\Delta T_{{\rm FWHM}} \Delta S_{M}^{max} = 100 \times 3 \approx 300$~J/Kg.  This value is comparable to the best known MCE materials. 

When the temperature is lowered below $50$~K, -$\Delta S_M$ becomes negative for small values of field. This behavior is known as the inverse Magneto Caloric Effect (MCE) and it indicates the presence of an antiferromagnetic component in the magnetic order \cite{Krenke2005a}.  With further increase in applied magnetic field, inverse MCE becomes less pronounced and finally disappears for $H\sim 5.5$ T.  When $H>6$ T, -$\Delta S_M$ approaches a positive value and magnitude increases with increasing magnetic field.  At $H=8$ T, -$\Delta S_M$ attains a value of $2.7$~J/kg-K which is comparable to -$\Delta S_M$ value observed near $T_{\mathrm C} \approx 195$~K\@.  The change in sign of -$\Delta S_M$ at low temperature and a downturn in $M(T)$ is indicative of another magnetic transition, possibly antiferromagnetic.   


{\it Heisenberg Model for Nd$_{2}$NiMnO$_{6}$:}
The experimental data suggests that {Nd$_{2}$NiMnO$_{6}$} is a promising candidate for application in magnetocaloric based cooling technology. Additionally, the results are indicative of the presence of different magnetic phases as a function of temperature. 
In order to better understand the experimental results, we propose a classical Heisenberg model on a body centred cubic lattice as the simplest phenomenological model for magnetism in {Nd$_{2}$NiMnO$_{6}$}. The Hamiltonian is given by, 
\vspace{0.3cm}
\begin{eqnarray}
{\cal H} &=& J_{1} \sum \limits_{nn}  {\bf S}_{Mn} \cdot{\bf S}_{Ni} + J_{2} \sum \limits_{nn} {\bf S}_{Nd} \cdot{\bf S}_{Mn} \nonumber \\
&& + J'_{2} \sum \limits_{nn}{\bf S}_{Nd} \cdot{\bf S}_{Ni}- H \sum_{i}  {S_{\it iz}}.  \label{heisen-perov}
\end{eqnarray}
\noindent
\\
The summation in the first three terms is over nearest neighbors ($nn$) and that in the last term is over all sites. 
We take $J_{1}$ as the ferromagnetic coupling between magnetic moments residing on Mn and Ni. $J_{2}$ ($J'_{2}$) is the weak antiferromagnetic coupling of Nd to Mn(Ni) moments. $H$ denotes the strength of uniform external magnetic field.  Justification for this choice of exchange interactions comes from the fact that a ferromagnetic transition is observed at a higher temperature while signatures of antiferromagnetic correlations are evident at lower temperatures.  The ferromagnetic exchange most likely arises due to double exchange between the two transition metal ions Mn$^{4+}$ and Ni$^{2+}$ having different valence states.  The antiferromagnetic exchange between the lanthanide Nd$^{3+}$ and the transition metal could arise from a super-exchange mechanism.  
Assuming the oxidation states of Mn$^{4+}$, Ni$^{2+}$ and Nd$^{3+}$, we take $S_{Mn} = S_{Nd} = 3/2$ and $S_{Ni} = 1$.

Although $J_2$ and $J'_2$ are not equal in general, given that $J_2 (J'_2) < |J_1|$ we assume $J_2 = J'_2$ for simplicity.
We set $J_{1}=-1$ and keep $J_{2}$ as a tunable parameter of the model which can be fixed by comparing simulations with the experimental results.  All energy scales are expressed in terms of $|J_{1}|$. To study the  finite temperature behavior of this model we employ the standard Markov chain Monte Carlo method with Metropolis algorithm. Simulations are carried out on lattices with $2 \times 16^{3}$ sites. 

In order to characterize and understand the various magnetic orderings, we compute component resolved sublattice magnetizations, given by,
\begin{equation}
M^{\mu}_{A(B)} = \frac{1}{N}  \left \langle \sum_{i \in A(B)} {S}_{i\mu}  \right \rangle,
\end{equation}
where $N$ is the total number of spins and the angular bracket denotes the thermal average over Monte Carlo generated equilibrium configurations.
$A$($B$) denotes the magnetic sublattice consisting of A(B and B$'$) sites in the DP structure. The total magnetization can then be obtained
as $M^{\mu} = M^{\mu}_A + M^{\mu}_B$. 
The strength of the MCE is estimated by computing the change in magnetic entropy as a function of temperature at different values of applied magnetic fields.
The change in magnetic entropy $(\Delta S_{M})$ is given by, 
\begin{equation}\label{entropy}
\Delta S_{M}(T_{i},H) = \sum \limits_{j=1}^{p} \frac{M^{z}(T_{i+1},H_{j})- M^{z}(T_{i-1},H_{j})}{T_{i+1}-T_{i-1}} (H_{j+1}-H_{j}).
\end{equation} 
Eq.~(\ref{entropy}) above is the discrete version of the continuum definition of $\Delta S_{M}$ given in Eq.~(\ref{Maxwell_eqn}). $\Delta S_{M}(T_{i},H)$ is thus the change in magnetic entropy at a specific value $T_{i}$ of temperature. $H_{j}$ is a result of the uniform discretization of interval $[0,H]$ such that, $H_1=0$ and $H_p=H$. 

\begin{figure}[t]
\includegraphics[width=0.9\columnwidth,angle=0, clip = 'True' ]{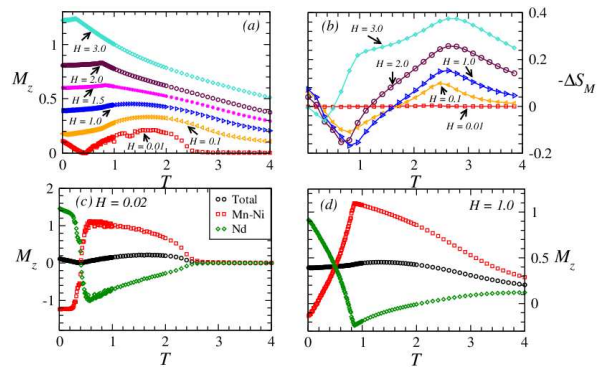}
 \caption{(Color online) ($a$) $z$ component of total Magnetization ($M_{z}$) as function of temperature ($T$) at various strengths of external field, ($b$) temperature variation of change in magnetic entropy $(\Delta S_{M})$ under external fields, ($c$) and ($d$) temperature dependence of $M_{z}$ for {Mn}-{Ni} and {Nd}-sublattice and total $M_{z}$ for two sublattices at $H=0.02$ and $H=1.0$. Results are obtained on $2 \times 16^{3}$ lattice with $J_{1}=-1.0$ and $J_{2}=0.17$.} 
 \label{fig46}
\end{figure}

{\it Simulation results and discussions:}
We now present the results for Hamiltonian Eq. (\ref{heisen-perov}) obtained using Monte Carlo simulations. At low strengths of magnetic field, the magnetization $M_z(T)$ begins to increase upon lowering the temperature below a critical value, indicating the presence of a paramagnetic to ferromagnetic transition at $T_c \simeq 2.5$ (Fig. \ref{fig46} ($a$)). The magnetization then shows a dip and rise feature at lower temperatures, which is a consequence of the antiferromagnetic coupling between $A$ and $B$ sublattices. With increasing magnetic field, the ferromagnetic transition broadens, as expected, and the dip evolves into a cusp-like feature. To gain further insight into the evolution of $M_z(T)$, we plot $M_{z}$ for the two sublattices separately. For $H=0.02$, the Ni-Mn sublattice begins to develop spontaneous magnetization in the direction of field. The Nd sublattice, which is antiferromagnetically coupled to the Ni-Mn sublattice, begins to have an induced order opposite to the applied field (see Fig. \ref{fig46} ($c$)). Since the local moment on $A$ sublattice is larger than that on $B$ sublattice, the ordered moment on $A$-sublattice exceeds that on $B$-sublattice at low temperature. This leads to a flipping of the orientation of the two sublattices, causing the cusp feature in magnetization. The ground state is a collinear ferrimagnetic state in which Nd moments align parallel to the magnetic field. For $H=1.0$ we notice from Fig. \ref{fig46} ($d$) that at high temperatures, Nd moments gain a small ferromagnetic moment in the direction of the field. However, the antiferromagnetic coupling between the two sublattices competes with this ferromagnetic state and at low temperature the $A$-sublattice again orders opposite to $B$-sublattice. As for low fields, upon lowering the temperature further the ordered moments of the two sublattices swap directions when the magnetic moment of $A$-sublattice becomes more than that of $B$-sublattice. 

\begin{figure}
 \includegraphics[width=0.6\columnwidth,angle=0, clip = 'True' ]{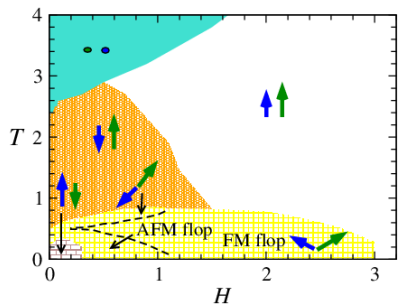}
 \caption{(Color online) $H$-$T$ phase diagram obtained with Monte-Carlo simulations. Blue arrow represents the magnetic moment of Nd-sublattice and the green arrow represents the magnetic moment of Mn-Ni sublattice. The blue and green dot depicts paramagnetic ordering for Nd-sublattice and Mn-Ni sublattice respectively.} 
 \label{fig48}
\end{figure}

To further understand the nature of magnetic states we also track the other two components of magnetization, $i.e$, $M_{x}$, $M_{y}$ 
(see supplementary figure). 
For large fields, finite values of $M_{x}$ and  $M_{y}$ for both sublattices indicate the presence of a spin-flop state.
Given that the antiferromagnetic correlations among all components of magnetization on two sublattices are retained, we name this state as AFM-flop state.  For still higher magnetic fields, the ground state becomes a FM-flop state, characterized by finite $M_x$ and $M_y$ on both sublattices and a ferromagnetic correlation between $z$-components of the sublattice magnetizations 
These results are summarized collectively as a phase diagram in Fig. \ref{fig48}. 
 
 
Next, we explicitly estimate the magnitude of the magnetocaloric effect by calculating -$\Delta S_{M}$ in our Monte-Carlo simulations using Eq. (\ref{heisen-perov}). The results are shown in Fig. \ref{fig46} (b). We find a peak and a dip at low magnetic field strengths corresponding to the two magnetic transitions. The ferromagnetic transition is characterized by -$\Delta S_{M}>0$ and the antiferromagnetic transition by -$\Delta S_{M}<0$. At higher fields the antiferromagnetic order is suppressed, and we find two peak structure which eventually evolves into one broad peak for further high magnetic field strengths. This broad peak leads to an enhanced working range in temperature for the magnetocaloric effect. The evolution of -$\Delta S_{M}<0$ calculated via Monte Carlo is qualitatively in agreement with that observed in experiment. 

\vspace{0.5cm}
{\it Conclusions:}
We find that the $T_C$ for Nd$_{2}$NiMnO$_{6}$ is dramatically enhanced compared to Y$_{2}$NiMnO$_{6}$. The paramagnetic background of Nd interacting antiferromagnetically with the Ni-Mn sublattice leads to additional broadening of $\Delta S_M (T)$ profile resulting in a large RCP $\approx 300$~J/Kg.  Monte Carlo simulations of a phenomenological Heisenberg model reproduce the key features of the experiments. Furthermore, a simple picture in terms of two magnetic sublattices emerges. Tracking the behaviour of each sublattice allows us to clarify the basic origin of the effects observed in experiments. We confirm that the $A$ sublattice provides a paramagnetic background interacting with the $B$ sublattice leading to a broad $\Delta S_M (T)$ enhancing the temperature range of MCE applicability to $T_C-50~K < T < T_C+50~K$.  With the high $T_C \approx 195$~K, this means that Nd$_2$NiMnO$_6$ could be useful for MCE cooling close to ambient temperatures. The low-temperature magnetic phase diagram obtained theoretically is interesting and consists of collinear and non-collinear magnetic phases which need to be confirmed by experiments on single crystals. 

{\it Acknowledgments:}
We thank the X-ray facility at IISER Mohali.  YS acknowledges DST, India for support through Ramanujan Grant \#SR/S2/RJN-76/2010 and through DST grant \#SB/S2/CMP-001/2013.
We acknowledge the use of High Performance Computing Facility at IISER Mohali. 
K.P. acknowledges support via CSIR/UGC fellowship.


%

\end{document}